\documentclass[preprint, amsfonts,showpacs,longbibliography,citeautoscript,superscriptaddress]{revtex4-2}
\usepackage[toc,page]{appendix}
\usepackage[british]{babel}
\usepackage[utf8]{inputenc}
\usepackage{amsmath,amssymb}
\usepackage{graphicx}
\usepackage{ae}
\usepackage{esdiff}
\usepackage{braket}
\usepackage{bbm}
\usepackage{simplewick}
\usepackage{float}
\usepackage{braket}
\usepackage{simplewick}
\usepackage{color}
\usepackage{framed}
\usepackage[caption=false]{subfig}
\usepackage[pdftex,colorlinks=true,linkcolor=blue,citecolor=blue,urlcolor=black]{hyperref}
\usepackage{hyperref}
\usepackage{mathtools}
\usepackage{enumitem}
\usepackage{ragged2e}
\usepackage{floatflt}
\usepackage{import}
\usepackage{titlesec}
\usepackage{multirow}
\usepackage{etoolbox}
\usepackage{appendix}
\usepackage{dsfont}
\usepackage{comment}
\usepackage{xcolor}
\usepackage{blindtext}
\usepackage[makeroom]{cancel}
\usepackage{mathtools}
\usepackage{lineno}

\begin{document}

\newcommand{\norm}[1]{\left\lVert#1\right\rVert}

\def \r{{\boldsymbol{r}}}
\def \k{{\boldsymbol{k}}}
\def \p{{\boldsymbol{p}}}
\def \q{{\boldsymbol{q}}}
\def \dl{\frac{\partial}{\partial l}}
\def \P{{\boldsymbol{P}}}
\def \K{{\boldsymbol{K}}}
\def \piph{\Pi_\text{ph}}
\def \sign{ \text{sign}}
\def \lamt{\tilde{\lambda}}
\def \intk{{\int_\textbf{k}}}
\def \Ims{\text{Im} [ \Sigma^R(\omega) ]}
\def \Gammabs{\Gamma_{\text{bs}}}
\def \aq{|\q|}
\def \ak{|\k|}
\def \Vt{\tilde V}
\def \omegap{\omega^\prime}

\definecolor{mgrey}{RGB}{63,63,63}
\definecolor{mred}{RGB}{235,97,51}
\newcommand{\mg}[1]{{\color{mgrey}{#1}}}
\newcommand{\mr}[1]{{\color{mred}{#1}}}

\newcommand{\red}[1]{{\color{red}{#1}}}
\newcommand{\blue}[1]{{\color{blue}{#1}}}
\newcommand{\al}[1]{ \begin{linenomath*} \begin{align}#1\end{align} \end{linenomath*}}
\newcommand{\beq} {\begin{equation}}
\newcommand{\eeq} {\end{equation}}
\newcommand{\bea} {\begin{eqnarray}}
\newcommand{\eea} {\end{eqnarray}}
\newcommand{\be} {\begin{equation}}
\newcommand{\ee} {\end{equation}}

\def\BigColSep{\setlength{\arraycolsep}{50pt}}

\title{Quantum phase transition in a clean superconductor with repulsive dynamical interaction}

\author{Dimitri Pimenov}\email{dpimenov@umn.edu}
\author{Andrey V.\ Chubukov}
\affiliation{William I. Fine Theoretical Physics Institute, University of Minnesota, Minneapolis, MN 55455, USA}

\begin{abstract}
{\bf{We consider a model of electrons at zero temperature, with a repulsive interaction which is a function of the energy transfer. Such an interaction can arise from the combination of electron-electron repulsion at high energies and the weaker electron-phonon attraction at low energies. As shown in previous works,  superconductivity can develop despite the overall repulsion due to the energy dependence of the interaction, but the gap $\Delta(\omega)$ must change sign at some (imaginary) frequency $\omega_0$ to counteract the repulsion. However, when the constant repulsive part of the interaction is increased, a quantum phase transition towards the normal state occurs. We show that, as the phase transition is approached, $\Delta$ and $\omega_0$ must vanish in a correlated way such that $1/|\log[\Delta(0)]| \sim \omega_0^2$. We discuss the behavior of phase fluctuations near this transition and show that the correlation between $\Delta(0)$ and $\omega_0$ locks the phase stiffness to  a non-zero value.}}
\end{abstract}
\maketitle

\section*{\uppercase{Introduction}}

Understanding  the nature of the ``pairing glue'', which enables Cooper pair formation of fermions, is one of the key steps towards a comprehensive scenario of superconductivity for a given material.  In strongly correlated materials, like cuprates, iron-based, heavy-fermion, and organic materials, the attractive pairing interaction is likely of electronic origin. Near a quantum phase transition, such an attraction often takes a
 more concrete form of an effective four-fermion interaction, mediated by soft collective fluctuations of the corresponding  order-parameter. Most often, the attraction emerges in a channel different from an ordinary $s$-wave, in which case the superconductivity is labeled as an unconventional one.

For more conventional metals the symmetry of the pairing gap is $s$-wave, and the attraction
 is believed to come from electron-phonon interaction. This is the backbone of the ``conventional" BCS theory of superconductivity. Still, to fully understand the phononic mechanism of $s$-wave superconductivity, one must explain why it is not overshadowed by the Coulomb repulsion, which is seemingly much larger. The frequently cited explanation
  \cite{tolmachev1958new, https://doi.org/10.1002/prop.19580061102,PhysRev.167.331,PhysRev.148.263,PhysRev.125.1263,RevModPhys.62.1027} is that the repulsive Coulomb repulsion is logarithmically renormalized down between the Fermi energy $E_F$ and the Debye energy $\Omega_D$ (the Tyablikov-McMillan logarithm), and at energies below $\Omega_D$ becomes smaller than the electron-phonon attraction, if the ratio $E_F/\Omega_D$ is large enough.

Upon closer examination, this explanation appears somewhat incomplete as Tyablikov-McMillan renormalization holds for the full interaction, i.e., for the sum of electron-electron and electron-phonon interactions, and under the renormalization this full interaction decreases, but does not change sign. It has been realized by several authors
\cite{gurevich1962possibility, PhysRev.148.263, PhysRevB.28.5100, PhysRevB.94.224515, PhysRevB.96.235107, PhysRevB.98.104505, PhysRevB.100.064513} that the underlying reason why electron-phonon superconductivity holds despite larger Coulomb interaction, is that the full interaction on the Matsubara axis (where it is real) is a dynamical one, $V(\Omega_m)$, and although a  phonon-mediated attraction does not invert the sign of $V(\Omega_m)$, it nevertheless reduces it at frequencies below the Debye energy. It was argued that an ``average" repulsive  $V(\Omega_m)$ can be effectively eliminated from the equation for the pairing gap $\Delta (\omega_m)$, by choosing a solution which changes  sign as a function of $\omega_m$. This bears some similarity to how, for an electronic pairing, a static Coulomb repulsion is effectively eliminated by choosing a sign-changing, non-$s$-wave spatial structure of the gap function.

A convenient way to model the dynamical $V(\Omega_m)$, suggested in Refs.~\cite{PhysRevB.100.064513,PhysRevB.28.5100,PhysRevB.94.224515,PhysRevB.96.235107, PhysRevB.98.104505},   is to treat it as a sum of  two parts: a constant repulsive part of strength $f$, representing the renormalized instantaneous Coulomb repulsion, and a frequency-dependent attractive part, due to electron-phonon interaction:
\beq
V(\Omega_m) \propto  f - \frac{1}{1+ (\Omega_m/\Omega_1)^2}\ ,
\label{th_1}
\eeq
 where $\Omega_1$ is of order of the Debye energy.  A similar reasoning has been applied~\cite{PhysRevB.96.235107} to dynamically screened electron-electron interaction, where $\Omega_1$ is of the order of plasma frequency.

 For $f >1$,   $V(\Omega_m) > 0$ for all frequencies, yet for $1<f < f_c$,  superconductivity emerges below a finite $T_c$, which contains $f$ in the combination $f/[1 + \text{const.}\times f \log(E_F/\Omega_1)]$.  For a given $f$ and large enough $\log(E_F/\Omega_1)$, the Coulomb repulsion becomes logarithmically small, and one recovers the McMillan formula for $T_c$. One the other hand, at a given  $E_F/\Omega_1$, at large enough $f > f_c$, the repulsion becomes too strong and superconductivity vanishes. Obviously, $T_c$ and the magnitude of the gap $\Delta (\omega_m)$ vanish at $f=f_c$.

   It is the goal of the present work to understand the nature of the $T=0$ quantum phase transition between a superconducting state at  $f <f_c$ and a normal state at $f > f_c$.
    Specifically, we resolve   the following puzzle: on the one hand, the gap $\Delta (\omega_m)$ must change sign  at some finite $\omega_m = \omega_0$, otherwise there would be no solution of the gap equation for $f >1$.  On the other hand,  for any finite $\omega_0$, $\Delta (0)$ is non-zero, in which case
      the linearized gap equation does not have a solution as the pairing kernel contains an infrared-divergent Cooper logarithm, which is not regularized  at $T = 0$
       and therefore does not admit a solution.
        We argue analytically and check numerically that as $f$ approaches $f_c$ from below,      $\omega_0 \rightarrow 0$ and $\Delta \rightarrow 0$ in tune with each other, such that $1/|\log \Delta| \sim \omega_0^2$.

We also analyze the spectrum of gapless phase fluctuations near $f = f_c$.  We show that because of the relation between $\Delta$ and $\omega_0$, the superfluid stiffness  remains finite as $f$ approaches $f_c$ from below. This is in marked contrast with the behavior of the stiffness near the end point of superconductivity at $T=0$ in a system with   magnetic impurities (Abrikosov-Gorkov theory, \cite{RevModPhys.78.373, abrikosov1959theory, abrikosov1959superconducting, PhysRev.136.A1500}).
{In this situation, the destruction of superconductivity occurs via pair-breaking due to the impurity-induced self-energy, and } the superfluid stiffness gradually vanishes as the
    system approaches the $T=0$ phase transition.

That  superconductivity  vanishes when  $\omega_0 =0$  can also be interpreted from a topological viewpoint, because $\omega_m = \omega_0$ is a center of a dynamical vortex: the anti-clockwise circulation of the phase of
           $\Delta (z)$, $z = \omega' + i \omega^{''}$, around this point is $2\pi$, Refs.\ \cite{PhysRevB.104.L140501, PhysRevB.103.024522, PhysRevB.103.184508}.
            There is no way to eliminate this dynamical vortex   as there are no anti-vortices in the upper half-plane of frequency  (their presence would be incompatible with the analyticity of $\Delta (z)$). Hence,  as long as superconducting order is present, $\omega_0$ must remain finite.  The only possibility for a vortex to disappear without destroying superconductivity is when it moves to an infinite frequency.
              For the model of Eq. (\ref{th_1}) this holds at $f =0$, and for $f <0$ the gap function $\Delta (\omega_m)$ on the Matsubara axis is nodeless.

{ We note in passing
that
   a vortex on the Matsubara axis gives rise to $2\pi$  winding of the phase of $\Delta (\omega)$ on the real frequency axis, between $\omega = -\infty$ and $\omega = \infty$.
   Such phase winding
    necessary leads to nodes in the real and imaginary parts of $\Delta(\omega)$, which can be detected by spectroscopic experiments, e.g.\ ARPES \cite{Damascelli_2004}. }

\section*{\bf{\uppercase{Results}}}

\paragraph*{\bf{Model}.}\label{modelsec}

We consider a spatially isotropic model of interacting spin-$1/2$ fermions at zero temperature in $d$ dimensions, described by the effective low-energy action
\begin{linenomath*}
 \begin{linenomath*} \begin{align}
&\mathcal{S} = \sum_{\sigma} \int_k \bar{\psi}_\sigma(k)(i\omega - \xi_\k) \psi_\sigma(k)\ \\ & \notag  +  
\int_{k,k^\prime,q}  V(\omega - \omega^\prime) \times   \bar{\psi}_\uparrow(k^\prime + q/2) \bar{\psi}_\downarrow(-k^\prime + q/2) \psi_\downarrow(-k + q/2) \psi_\uparrow(k+q/2) \ ,  \\ \notag
& \int_k = \int_{-\Lambda}^\Lambda \frac{d\omega}{2\pi} \int \frac{d\k}{(2\pi)^d}  \ ,
\end{align} \end{linenomath*}
\end{linenomath*}
 where $\Lambda$ is a UV cutoff of order $E_F$  and $\omega$ are Matsubara frequencies (here and below we label  Matsubara frequency as $\omega$ without subscript $m$).
The interaction $ V(\Omega)$ is taken to be
 a function of the energy transfer, but independent of momenta.
 We follow  earlier works \cite{PhysRev.99.1140, PhysRevB.100.064513,PhysRevB.94.224515, PhysRevB.96.235107, PhysRevB.98.104505} and
  set $V(\Omega)$ to
   \begin{linenomath*} \begin{align}
\label{Vtilde}
 V(\Omega) = \frac{2}{\rho} \times \tilde {V}(\Omega), \quad \tilde{V}(\Omega) =  \lambda \left(f - \frac{1}{1+\left(\Omega/\Omega_1\right)^2} \right) ,
\end{align} \end{linenomath*}
where $\rho$ is the single-spin density of states at the Fermi surface,  and
  $\Omega_1$ is of the order of Debye energy for the electron-phonon case
  (the factor of 2 is introduced for notational convenience).

\begin{figure}
\centering
\includegraphics[width=.6\columnwidth]{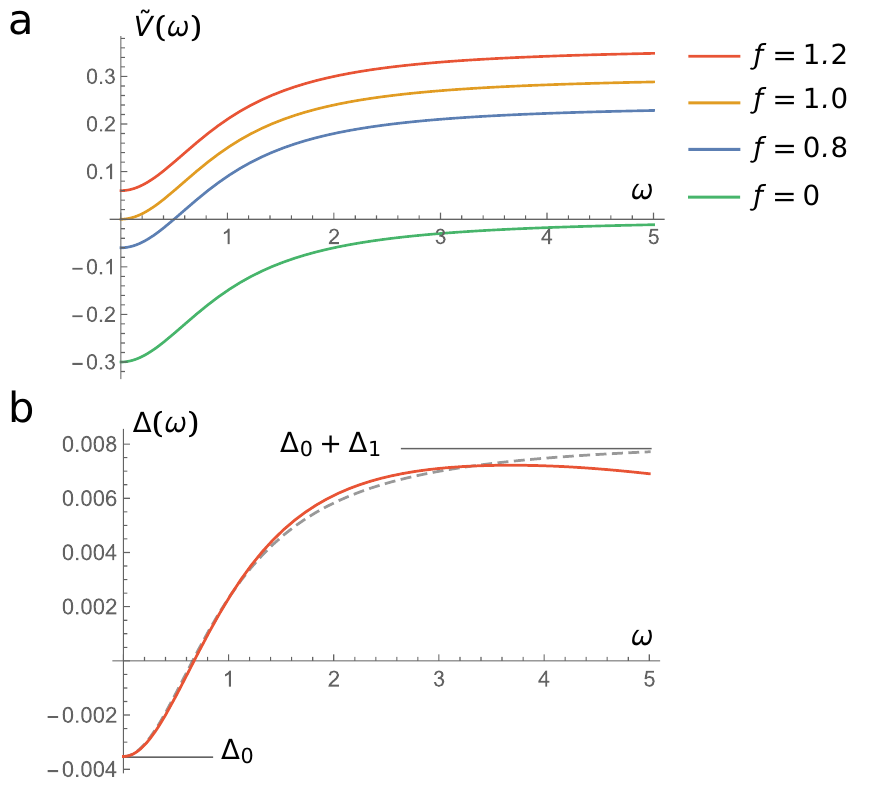}
\caption{\textbf{Dynamical interaction and typical gap function}. \textbf{a} $\Vt(\omega)$ for $\Lambda = 5, \lambda = 0.3$ and four different values of $f$.  \textbf{b} Numerical solution of the gap equation for $f = 1.2$. Gray dashed line corresponds to a fit to $\Delta(\omega)$ using the ansatz \eqref{bestfit}. }
\label{Vdelta1fig}
\end{figure}

In the following we measure all energies in units of $\Omega_1$, and hence set   $\Omega_1 = 1$  in Eq.~(\ref{Vtilde}). Then, $\Lambda \gg 1$. A discussion of the opposite low-density limit where $E_F, \Lambda \ll1 $ can be found in Ref.\ \cite{phan2021effect}.  For the known  physical realizations of Eq.~(\ref{Vtilde}),
      $f >1$, hence $V(\Omega)$ remains positive (repulsive) at all frequencies. For completeness, here we
       consider arbitrary $f$, but our key focus will still be on $f >1$.
     For  a generic $f$,  $V(\Omega)$ is purely attractive for $f \leq 0$, is attractive at small frequencies and repulsive at large frequencies for $0<f <1$, and is purely repulsive for $f \geq 1$,  see Fig.\ \ref{Vdelta1fig}a. The dimensionless $\lambda$ parametrizes the overall strength of the interaction. We assume $\lambda \leq 1$, this will allow us to neglect, at least qualitatively, the normal fermionic self-energy: {One can show that the leading self-energy effect is a mere renormalization of the coupling constant $\lambda \rightarrow \lambda/(1 + 2\lambda)$.}

\paragraph*{\bf{Gap equation.}}
\label{sec:gap}

To describe superconductivity, we perform a Hubbard-Stratonovich transformation in the spin-singlet, $s$-wave pa
iring channel and use a saddle point approximation. This procedure leads to the conventional
Eliashberg equation for the gap function~\cite{eliashberg1960interactions}, though without the additional contribution from the self-energy. On the Matsubara axis we have
\begin{linenomath*}
 \begin{linenomath*} \begin{align}
\label{Gapeq}
\Delta(\omega) &= - \int_{-\Lambda}^\Lambda d\omega^\prime \frac{\Delta(\omega^\prime) \Vt(\omega - \omega^\prime)}{\sqrt{(\omega^\prime)^2 + |\Delta(\omega^\prime)|^2}}  \ .
\end{align} \end{linenomath*}
\end{linenomath*}
The interaction $ \Vt(\omega - \omega^\prime)$ is real on the Matsubara axis, which allows us to set $\Delta (\omega)$ to be real by properly choosing its phase.  At the same time, because the interaction is a function of the frequency transfer, one can search for even-frequency and odd-frequency $\Delta (\omega)$.  In this communication, we focus on the even-frequency solutions.
For even frequency $\Delta (\omega) = \Delta (-\omega)$, the gap equation can be rewritten as
 \begin{linenomath*} \begin{align}
\label{symgapeq}
\Delta(\omega) = -\int_{0}^\Lambda d\omega^\prime \frac{\Delta(\omega^\prime) \left[ \tilde V(\omega - \omega^\prime)+ \tilde V(\omega + \omega^\prime)\right]}{\sqrt{(\omega^\prime)^2 + \Delta(\omega^\prime)^2}}\  .
\end{align} \end{linenomath*}
 It is obvious that for $f >1$, when  $\Vt> 0$,  $\Delta(\omega)$ must change sign at some frequency
   $\omega_0$  as otherwise
 the left hand side and the right hand side of Eq.\ \eqref{symgapeq} would have opposite signs. The value of $\omega_0$
  is chosen to minimize the effect of a repulsive $f$ in Eq. (\ref{Vtilde}). This has been discussed before~\cite{https://doi.org/10.1002/prop.19580061102, PhysRev.167.331,PhysRev.148.263,PhysRev.125.1263,RevModPhys.62.1027,PhysRevB.100.064513,PhysRevB.28.5100,PhysRevB.98.104505} and we just state the results.  First, $\omega_0$ is finite for all $f >0$ if $\Lambda < \infty$. If $\Lambda$ is finite, $\Delta (\omega)$ has a node as long as $\omega_0 < \Lambda$.
  Second, optimizing $\omega_0$ in the limit $\log(\Lambda) \gg 1$,
   one obtains that a repulsive $f$ effectively gets reduced to $f/[1 + \lambda f \log{(\Lambda)}] \to 1/\lambda \log(\Lambda)$. This gives rise to the McMillan formula for $T_c \propto e^{-1/(\lambda - \mu^*)}$, in which
   $\mu^* \approx 1/\log{\Lambda}$ is the contribution from the repulsion. Third,  for any  finite $\Lambda$, the repusive part of the interaction gets reduced, but cannot be completely eliminated. As a result, superconductivity exists at $f$
    smaller than some critical $f_c >1$.

In Fig.~\ref{Vdelta1fig}b  we present the numerical solution of the non-linear gap equation (\ref{symgapeq}) for
 some representative  $\Lambda$ and $1< f < f_c$. We clearly see that $\Delta (\omega)$ changes sign at some finite $\omega_0$.  It reaches a finite value at $\omega =0$ and then saturates at some other finite value, of opposite sign
     at $\omega_0 \ll \omega < \Lambda$. The numerical solution has been obtained by a ``damped iteration'' method, in which only a certain portion of  $\Delta (\omega)$ is updated at each step of iterations.  This method improves the convergence of the iteration procedure~\cite{PhysRevB.100.064513}.\\

\paragraph*{\bf{Quantum phase transition towards a superconductor with nodeless $\Delta (\omega)$.}}
\label{topQPTsec}
Before we proceed to the case $f \approx f_c$, we briefly discuss the transition towards the state with a nodeless $\Delta (\omega)$. As stated above, this transition occurs at $f =0$ when $\Lambda = \infty$, i.e., Eq.~(\ref{symgapeq}) holds at all frequencies. This transition can be classified as topological because it separates two states with and without a dynamical vortex. As $f$ is reduced towards $f=0$, $\omega_0$ increases, i.e., the core of the dynamical vortex successively moves to larger $\omega$. At $f=0$  it  reaches $\omega = \infty$ and disappears.

\begin{figure}
\centering
\includegraphics[width=.7\columnwidth]{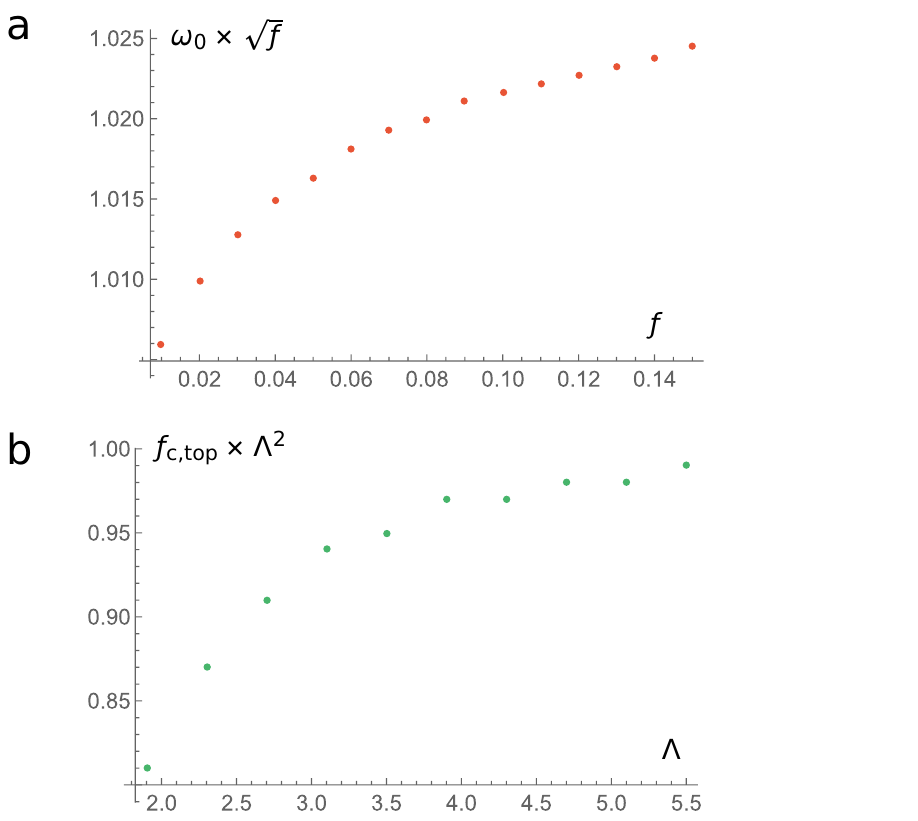}
\caption{\textbf{Scaling behavior at the transition towards the nodeless superconductor}. \textbf{a} Numerical check of Eq.\ \eqref{omega0sqrt} for small $f$. \textbf{b} Numerical check of Eq.\ \eqref{fcritLambda} for moderate $\Lambda$. The apparent oscillations are due to discretization of $f$ in the numerics. Used parameter: $\lambda = 0.1$. }
\label{topnodes}
\end{figure}

We now argue that the dependence of  $\omega_0$ of $f$ has a simple form
 \begin{linenomath*} \begin{align}
\label{omega0sqrt}
\omega_0 = f^{-1/2} \quad \text{as} \quad f \rightarrow 0 \ .
\end{align} \end{linenomath*}
This can be obtained as follows: Let $\Delta_a(\omega)$ be the solution of the gap equation at $f = 0$:
 \begin{linenomath*} \begin{align}
&\Delta_a(\omega) =  \lambda \int_0^\Lambda d\omega^\prime \left (\frac{1}{1+(\omega - \omega^\prime)^2} + \frac{1}{1 + (\omega \ {+} \ \omega^\prime)^2} \right)  \times \frac{\Delta_a(\omega^\prime)}{\sqrt{(\omega^\prime)^2 + \Delta_a(\omega^\prime)^2}} \ .
\label{Deltaa}
\end{align} \end{linenomath*}
Since $\Vt(\omega)$ is purely attractive for $f = 0$, $\Delta_a$ has a fixed sign.
 At large $\omega$,  $\Delta_a (\omega) = A/\omega^2$, where
 \begin{linenomath*} \begin{align}
\label{Deltaaquadratic}
 A =
  2 \lambda \int_0^\Lambda d\omega^\prime \frac{\Delta_a(\omega^\prime)}{\sqrt{(\omega^\prime)^2 + \Delta_a(\omega^\prime)^2}} \ .
\end{align} \end{linenomath*}
Now let $\Delta_b(\omega)$ be the solution of the gap equation at small but finite $f$.
 To the  leading order in $f$, we obtain
  \begin{linenomath*} \begin{align}
\label{Deltab}
\Delta_b(\omega) \simeq \Delta_a(\omega) - A f = A \left (\frac{1}{\omega^2} -f\right) \ .
\end{align} \end{linenomath*}
At the node, $\Delta_b(\omega_0) = 0$, hence   $\omega_0 = f^{-1/2}$. For large but finite $\Lambda$,
  the phase transition occurs when $\omega_0 = \Lambda$.  The corresponding critical $f$ for a topological transition  is then  \begin{linenomath*} \begin{align}
\label{fcritLambda}
f_{c,\text{top}}= \Lambda^{-2} \ .
\end{align} \end{linenomath*}
In Fig.~\ref{topnodes} we checked these results by solving the gap equation numerically. The agreement between the numerical and analytical results is perfect.\\

\paragraph*{\bf{Quantum phase transition towards normal state}.}
\label{normQPTsec}

We now consider the system behavior near the  $T=0$ transition towards the normal state at $f = f_c >1$, when the pairing interaction $V(\omega - \omega')$ is positive (repulsive) at all frequencies.  We assume and then verify that the transition is continuous, i.e., at $f = f_c -0^+$, $\Delta (\omega)$ is infinitesimally small. Like we said in the Introduction, to understand this transition one has to resolve the following puzzle:
if infinitesimally small $\Delta (\omega)$ tends to a finite value at $\omega =0$, like, e.g., in Fig.\ \ref{Vdelta1fig}b,  the right hand side of the linearized gap equation gives rise to a divergent Cooper logarithm. Because $T=0$, the logarithmical divergence is not cut. The only way to avoid this divergence is to place the node (i.e., the vortex core) right at $\omega = 0$. But then the gap  becomes sign-preserving at all finite $\omega$, and for such $\Delta (\omega)$ there is no  solution  of the gap equation for a
purely repulsive interaction.

As we now show, the resolution of this problem is to let both $\Delta$ and $\omega_0$ vanish in a correlated way as
$f \to f_c$ from below. To simplify the analysis, we first note that for all $f < f_c$, the gap function $\Delta (\omega)$ is well approximated by a simple form
 \begin{linenomath*} \begin{align}
\label{bestfit}
\Delta(\omega) = \Delta_0 + \Delta_1 \frac{\omega^2}{1+\omega^2} \ .
\end{align} \end{linenomath*}
 A comparison with the numerical solution of the gap equation shows that this form is near-perfect for $\omega < 1$ and matches the numerical results reasonably well  for $\omega > 1$, see Fig.\ \ref{Vdelta1fig}b. {Such agreement is sufficient to extract the leading behavior near the phase transition (see below).}  The coefficients $\Delta_0, \Delta_1$ can be determined by inserting the ansatz (\ref{bestfit})  into \eqref{symgapeq} and expanding up to second order in $\omega$. After a straightforward algebra we obtain  \begin{linenomath*} \begin{align}
\label{Delta0form}
&\Delta_0 = -2\lambda \int_0^\Lambda d\omegap \left(f - \frac{1}{1+(\omegap)^2} \right) \times K(\omegap)  \\ &
\label{Delta1form}
\Delta_1 = 2\lambda \int_0^\Lambda d\omegap \frac{\left(3 (\omegap)^2 -1  \right)}{(1+ (\omegap)^2)^3} \times K(\omegap)  \\
&K(\omegap) = \frac{\Delta_0 + \Delta_1 \frac{(\omegap)^2}{1+ (\omegap)^2} }{\sqrt{ (\omegap)^2 + \left(\Delta_0 + \Delta_1\frac{(\omegap)^2}{1+(\omegap)^2} \right)^2 }} \ .
\label{Kdef}
\end{align} \end{linenomath*}
In the limit $\Delta_0, \Delta_1 \rightarrow 0$, the equations simplify to
 \begin{linenomath*} \begin{align}
\label{Deltaeqs}
\Delta_1 &= \frac{\Delta_1 \lambda}{6} - 2\lambda \ell \Delta_0 \\ \notag
\Delta_0 &= \Delta_0 \left[ - 2\lambda f L - 2\lambda \ell (f-1) \right] + \Delta_1 ( - 2\lambda f L + \lambda) \ .
\end{align} \end{linenomath*}
 where
 \begin{linenomath*} \begin{align}
 \quad  L = \log(\Lambda), ~~\ell = \log(1/\Delta_0) \gg L
\label{th_4}
\end{align} \end{linenomath*}
The value of the critical $f_c$ can be determined by evaluating the determinant of the set
\eqref{Deltaeqs} in the limit $\ell \rightarrow \infty$. We obtain
 \begin{linenomath*} \begin{align}
\label{fcformula}
f_c  = \frac{1 - 7/6 \times \lambda}{1- \lambda/6 - 2L\lambda} \ .
\end{align} \end{linenomath*}

{The divergence of $f_c$ at a critical value of $\lambda L$, which is evident from Eq. (\ref{fcformula})
 is not an artefact of the approximation to $\Delta(\omega)$, as we have checked numerically. Rather, it implies that by properly placing $\omega_0$, one can completely eliminate a constant repulsion $f$  even when $f$ is large. A detailed analysis of this effect will be presented elsewhere (in preparation).}

  The general trend that $f_c$ increases with increasing $\Lambda$ is also in agreement with
McMillan reasoning that the Coulomb repulsion is suppressed at large $\Lambda$.  In the following, we focus on $\lambda L \ll 1$, in which case
   \begin{linenomath*} \begin{align}
f_c  = 1 + \lambda(2L - 1) + \mathcal{O}((\lambda L)^2)
\end{align} \end{linenomath*}

Evaluating the determinant again, but this time for a finite $\ell$, we obtain to leading order in $\lambda L$ and $f_c - f$:
 \begin{linenomath*} \begin{align}
\label{scalingformula1}
\ell \simeq \frac{1}{2\lambda (f_c -f )} \times \left( \frac{1}{1- 2\lambda L} \right)^2
\end{align} \end{linenomath*} Using (\ref{th_4}) we find that $\Delta_0 = \exp(-\ell)$ vanishes exponentially fast as $f \nearrow f_c$.

From the first  equation in \eqref{Deltaeqs} we obtain
\beq
\frac{\Delta_0}{\Delta_1} \simeq - \frac{1}{2 \lambda \ell} \simeq - (f_c-f) \left(1-2 \lambda L\right)^2 \ .
\label{th_3}
\eeq
 The ratio is negative (hence $\omega_0$ is finite)  and progressively decreases when $f$ approaches $f_c$.
  Substituting $\Delta_0/\Delta_1$  into (\ref{bestfit}), we obtain
  \begin{linenomath*} \begin{align}
  \label{scalingformula2}
\omega_0 \simeq \sqrt{f_c -f } \times  (1 - 2\lambda L) .
\end{align} \end{linenomath*}
 We see that $\omega_0$ vanishes as $\sqrt{f_c -f }$, i.e., much more gradually than $\Delta_0$.

\begin{figure}
\centering
\includegraphics[width=.7\columnwidth]{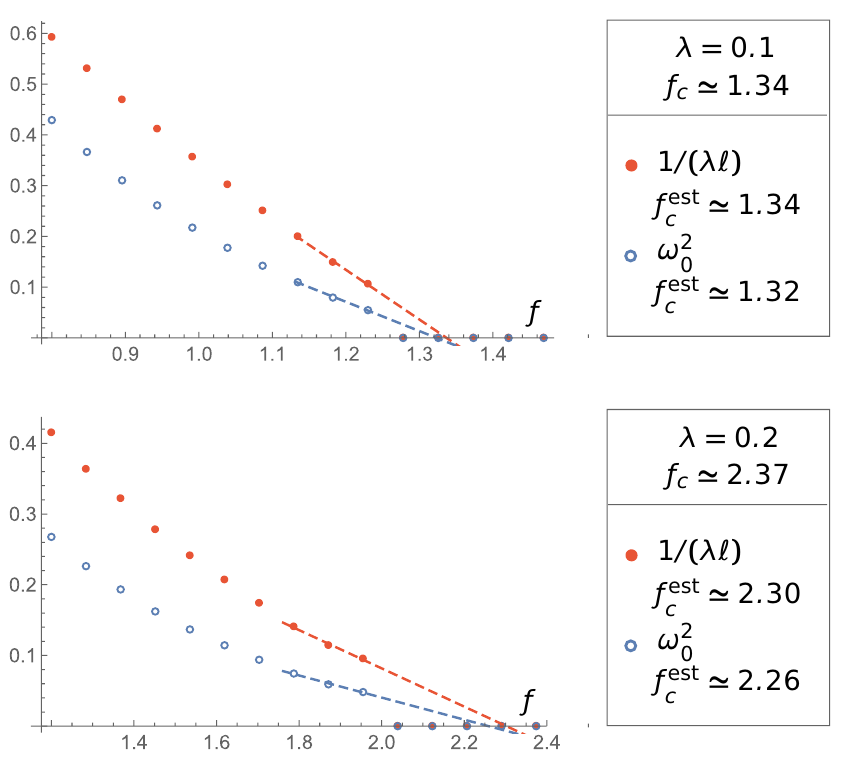}
\caption{\textbf{Scaling behavior at the transition towards the normal state}. Evolution of $1/(\lambda\ell)$ (filled circles) and $\omega_0^2$ (empty circles) close to the phase transition; both quantities vanish with approximately constant slope (i.e., are $\propto f_c-f)$, as expected from Eqs.\ \eqref{scalingformula1}, \eqref{scalingformula2}. Values $f_c^{\text{est}}$ shown in the plot legend are derived from linear extrapolation of the last three data points (dashed lines), showing semi-quantitative agreement with $f_c$ from Eq.\ \eqref{fcformula}.}
\label{scalingFig}
\end{figure}

 In Fig.~\ref{scalingFig} we verify the scaling forms of $\Delta_0$ and $\omega_0$ by extracting these two quantities from the numerical solution of the gap equation. The agreement between analytical and numerical results is quite good.\\

\paragraph*{\bf{Phase fluctuations near critical $f_c$.}}
\label{phaseflucsec}

For a more detailed characterization of the phase transition at $f=f_c$, we now look at
soft collective excitations in the system. These are  phase fluctuations, which in the absence of long-range Coulomb interaction are Goldstone modes of   the superconducting state.  Our goal is to derive the superfluid density and the dynamical compressibility, which enter the propagator of phase fluctuations, as functions of $q = (\Omega, \q)$, where ${\q}$ the total momentum of a Cooper pair and $\Omega$  is the total frequency.
 There are two ways to do this: either expand the action to second order in $\theta$ or analyze the pole  structure of the full particle-particle susceptibility at small $q$. These two methods yield consistent results; we will focus on the first one in the remainder of this section as we discuss the other one in the Methods section.

To obtain the propagator of low-energy phase fluctuations, we introduce the total momentum ${\q}$ and the total frequency $\Omega$ of a Cooper pair.  For convenience, we combine ${\q}$ and $\Omega$ into a ($d$+1)-dimensional variable $q = (\Omega, \q)$. In our mean-field solution the pairing involves fermions with frequencies $\omega$ and $-\omega$ and momenta ${\k}$ and $-{\k}$, i.e., $q$ is set to zero.  In other words, the mean-field gap is a function of $\omega$ but not of $q$. This mean-field solution corresponds to a minimum of the Luttinger-Ward functional. States away from the
 minimum are described by a fluctuating pairing field (order parameter) that depends on both $\omega$ and $q$. We illustrate this in Fig.\ \ref{pairingvariables}.

 Low-energy fluctuations around the mean-field solution correspond to slow variations of the phase of the order parameter
  $\theta (q)$:

\begin{figure}
\centering
\includegraphics[width=.7\columnwidth]{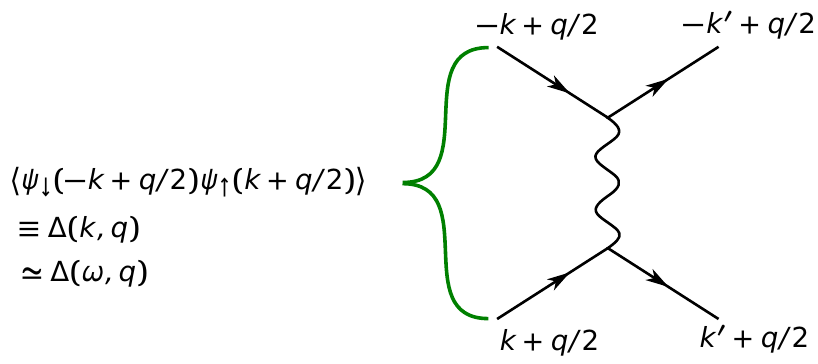}
\caption{\textbf{Energy-momentum dependence of the pairing field $\Delta$}. The brace represents the Hubbard-Stratonovich decoupling. Four-momenta $k = (\omega, \k), q = (\Omega, \q)$ are used. We always neglect the dependence of $\Delta$ on the relative momentum of a Cooper pair, $\k$. }
\label{pairingvariables}
\end{figure}

 \begin{linenomath*} \begin{align}
 \label{thetaapprox}
\Delta (\omega, q ) &\simeq  \Delta (\omega) \times \left[\delta^{(d+1)}(q)+ i\theta(q)\right]   \\
\bar{\Delta}(\omega, q) &\simeq \Delta (\omega) \times \left[\delta^{(d+1)}(q)- i\theta(-q)\right] \notag .
\end{align} \end{linenomath*}
 The expressions in the square brackets arise from small-$\theta$ expansion and Fourier transformation of the real-space phase factor
 \begin{linenomath*} \begin{align}
&\int \frac{d^{d+1}q}{(2\pi)^{d+1}} \exp[i\theta(x)] \exp[i (x\cdot q)] \simeq  \\ & \int \frac{d^{d+1}q}{(2\pi)^{d+1}} \left(1+ i\theta(x) \right) \exp[i (x\cdot q)]  = \delta^{(d+1)}(q)+ i\theta(q) \,
\notag
\end{align} \end{linenomath*}
where $x$ is the center-of-mass coordinate.

Inserting the expansion \eqref{thetaapprox} into the $\Delta$-dependent action, where the fermions have been integrated out,  and expanding to the second order in $\theta$, we obtain the following action for the $\theta$ field (see  Methods for details):
\begin{widetext}
 \begin{linenomath*} \begin{align}
\label{GSaction}
\mathcal{S}_\theta &= \int \frac{d\Omega}{2\pi} \frac{d\q}{(2\pi)^d} \theta(q) \theta(-q)
\left[ \int \frac{d\omega d\omega^\prime}{(2\pi)^2} \Delta(\omega) ( - V^{-1})(\omega - \omega^\prime) \Delta(\omega^\prime) - \Pi_\Delta(q) \right]  \ . \\
\label{Piqdefinition}
\Pi_\Delta(q) & = \int \frac{d\omega}{2\pi} \frac{d\k}{(2\pi)^d} \frac{\Delta^2(\omega)}{\omega_+^2 + E_+^2} \frac{1}{\omega_-^2 + E_{-}^2} \times \left[ \Delta_+\Delta_{-}  +\left( i\omega_+ + \xi_{+}\right)(i\omega_{-} + \xi_{-}) \right] , \\ \notag
\xi_{\pm} &= \xi_{\q/2 \pm \k}, \quad \omega_{\pm} = \Omega/2 \pm \omega , \quad \Delta_{\pm} = \Delta(\omega_{\pm}),  \quad E_{\pm} = \sqrt{ (\Delta_{\pm})^2 + (\xi_{\pm})^2} \ .
\end{align} \end{linenomath*}
\end{widetext}
Here,
 $V^{-1}(\omega - \omega^\prime)$ is the  inverse of $V(\omega - \omega^\prime)$ defined  by
 \begin{linenomath*} \begin{align}
\int \frac{d\omega_1}{2\pi}  V(\omega - \omega_1) V^{-1}(\omega_1 - \omega^\prime) = 2\pi \delta(\omega - \omega^\prime).
\end{align} \end{linenomath*}
{The expression for $\Pi_\Delta(q)$ in Eq.\ \eqref{GSaction} can directly be obtained from the expansion in $\theta$. Alternatively, it can be
obtained diagrammatically as a  particle-particle bubble  with form-factors $\Delta^2(\omega)$.
 (see Methods). }

 Multiplying both sides of Eq.\ \eqref{Gapeq} by $V^{-1}(\omega - \tilde{\omega}) \Delta(\tilde\omega)$ and integrating over $\omega, \tilde\omega$ we find that
 \bea
 &&\int \frac{d\omega d\omega^\prime}{(2\pi)^2} \Delta(\omega) ( - V^{-1})(\omega - \omega^\prime) \Delta(\omega^\prime) \nonumber \\
 && = \frac{\rho}{2} \int d\omega \frac{\Delta(\omega)}{\sqrt{\Delta(\omega)^2 + \omega^2}}
 \eea
 This expression  coincides with the particle-particle bubble in the limit of vanishing $q$:
  \begin{linenomath*} \begin{align}
\Pi_\Delta(0) &= \rho\int \frac{d\omega}{2\pi}  d\xi \frac{\Delta(\omega)^2}{\Delta(\omega)^2 + \xi^2 + \omega^2} \\\ &=
\frac{\rho}{2} \int d\omega \frac{\Delta(\omega)}{\sqrt{\Delta(\omega)^2 + \omega^2}} \ .  \notag
\end{align} \end{linenomath*}
As a result, the propagator of the field $\theta (q)$  is determined by $\Pi_\Delta(q) - \Pi_\Delta(0)$.  Expanding $\Pi_\Delta(q)$ to second order in $\Omega,\q$, we obtain
 \begin{linenomath*} \begin{align}
\label{GSactionfinal}
S_\theta  \simeq
\frac{\rho}{4} \int \frac{d\Omega}{2\pi} \frac{d\q}{(2\pi)^2} \theta(q) \theta(-q)
 \left[\frac{1}{d} n_s |v_F \q|^2 + \kappa \Omega^2 \right]\ .
\end{align} \end{linenomath*}
 The coefficient $n_s$ in the second line of \eqref{GSactionfinal} is the superfluid density, normalized by the density of electrons in the normal state.

   It  parametrizes the energy cost of spatial phase fluctuations of the order parameter and controls the supercurrent and magnetic response in the superconducting state.
   (The prefactor $1/d$ comes from averaging over $\cos[\measuredangle(\k,\q)]^2$ in $d$ dimensions).
   The factor  $\kappa$ is a dynamical compressibility which parametrizes the energy cost of temporal phase fluctuations. We find
\begin{widetext}
 \begin{linenomath*} \begin{align}
\label{nsintegral}
 n_s &=  \frac{1}{2} \int_{-\Lambda}^{\Lambda} d\omega  \frac{\Delta(\omega)^2}{\left(\Delta(\omega)^2 + \omega^2\right)^{3/2}} \\
 \label{kappaintegral}
\kappa &=\frac{1}{4} \int_{-\Lambda}^\Lambda d\omega \frac{\Delta(\omega)^2}{(\Delta(\omega)^2  + \omega^2)^{5/2}}  \times  \left[\Delta(\omega)^2 (3 - \Delta(\omega) \Delta(\omega)^{\prime\prime}) + \omega^2(-\Delta (\omega)\Delta^{\prime \prime}(\omega) + 3\Delta^\prime(\omega)^2 ) \right]    .
\end{align} \end{linenomath*}
\end{widetext}
 where the derivatives are with respect to $\omega$. A similar expression for $n_s$ was also obtained in Ref.\ \cite{doi:10.1143/JPSJ.80.044711}.

\begin{figure}
\centering
\includegraphics[width=.7\columnwidth]{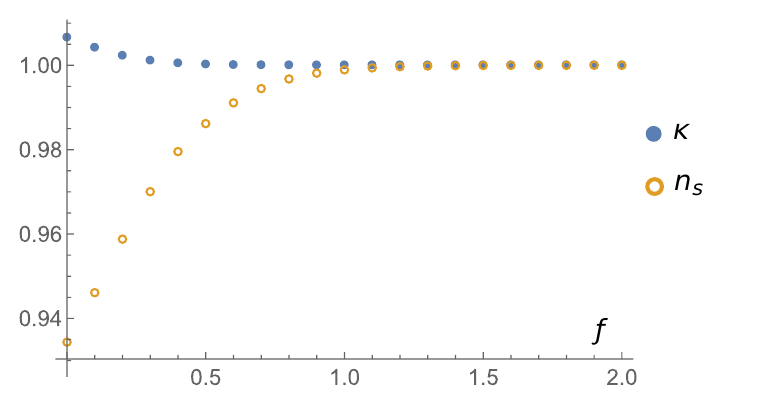}
\caption{\textbf{Normalized compressibility  and superfluid density}. $\kappa$ and $n_s$ are shown as  function of $f$. Used parameters: $\lambda = 0.3, \Lambda = 5$.}
\label{nskappaFig}
\end{figure}

For a weakly frequency dependent $\Delta (\omega)$, the expressions for $n_s$ and $\kappa$ are the same as in BCS theory, $n_s = \kappa =1$.  For a generic $\Delta (\omega)$, the BCS results are correct by order of magnitude, but the actual values of $n_s$ and $\kappa$ differ by BCS expressions by $O(1)$, see Fig. \ref{nskappaFig}.

   At $f \leq f_c$, $\Delta(\omega)$ is exponentially small, and the integrals are dominated by $\omega \lesssim \Delta(0)$.
    Because frequency variation of $\Delta (\omega)$ occurs at a much larger scale $\omega_0 \gg \Delta (0)$, the gap in (\ref{nsintegral}) and (\ref{kappaintegral})  can be approximated by a constant $\Delta (0)$. As a result, both $n_s$ and $\kappa$ tend to BCS values $n_s = \kappa =1$.  We verified this result in numerical calculations, see Fig.\ \ref{nskappaFig}.  The velocity of phase fluctuations also approaches the BCS value
 \begin{linenomath*} \begin{align}
v_s \rightarrow \frac{1}{\sqrt{d}} v_F \quad   \text{for} \quad  f \nearrow f_c \ .
\end{align} \end{linenomath*}
 At $f=f_c +0^+$,
  $n_s$ and $\kappa$ jump to zero.

We emphasize that the behavior of $n_s$ near the $T=0$ superconductor-normal state phase transition in a clean system at $f=f_c$  is different from the one at the $T=0$ superconductor-normal state phase transition  due to magnetic impurities. There, $n_s$ gradually vanishes
 at a critical impurity concentration
 {due to pair-breaking coming from the impurity-induced self-energy.
   As a result, }the penetration depth $\sim 1/\sqrt{n_s}$ diverges (Refs.~\cite{RevModPhys.78.373, abrikosov1959theory, PhysRev.136.A1500}).   At a technical level, this is because in the case of magnetic impurities the denominator of Eq.\ \eqref{nsintegral} contains an additional term proportional to the fermionic damping rate due to impurity scattering.  In our case, such a constant term is absent.  We expect, however, that it will appear if we extend the analysis of the superconductor-normal state phase transition  to a finite magnetic field $H$. We therefore expect that at a finite field, $n_s$  will vanish at critical $f_c (H)$.

\begin{figure}
\centering
\includegraphics[width=.7\columnwidth]{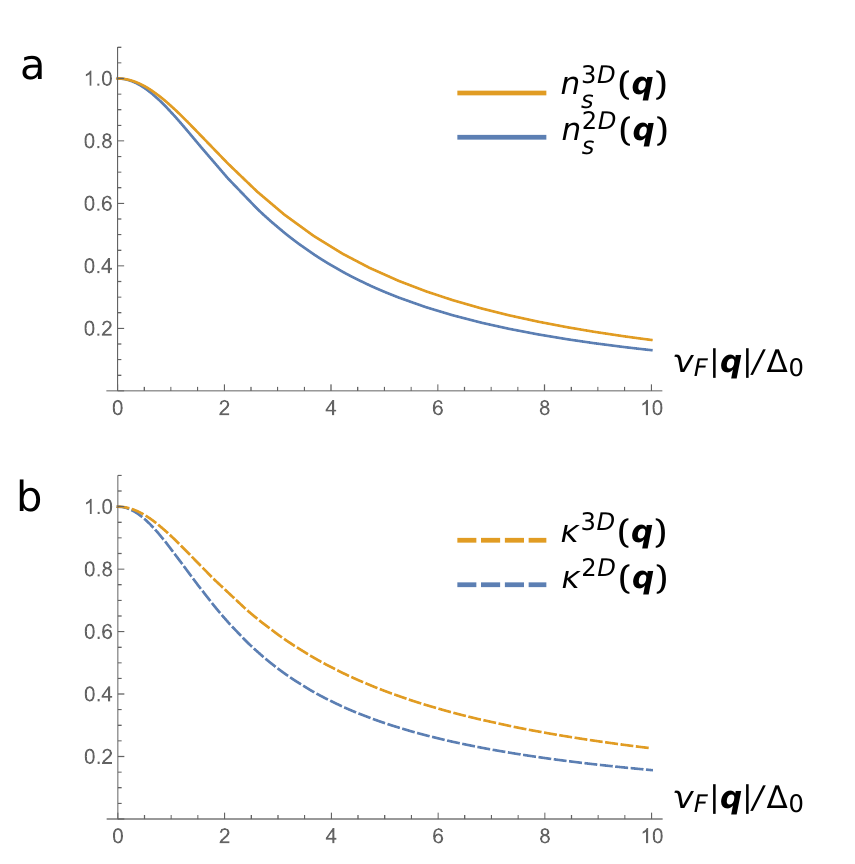}
\caption{\textbf{Generalized superfluid density}. \textbf{a}   $n_s(\q)$ determined from Eqs.\ \eqref{nsqanalytic}, \eqref{nsqanalytic3d} \textbf{b} $\kappa(\q)$, obtained in the same way as $n_s(\q)$.}
\label{nsqFig}
\end{figure}

Still, the discontinuity of $n_s$ at $f=f_c$ in our case holds only for superfluid density evaluated at zero momentum ${\q}$,  or, more accurately, at  $|v_F \q| \ll \Delta_0$.
 To analyze the behavior at larger $|{\q}|$,  we define a generalized momentum-dependent superfluid density as
 \begin{linenomath*} \begin{align}
&n_s(\q) \equiv \frac{\left( \Pi_{\Delta}(0) - \Pi_{\Delta}(\q,0) \right) }{\tfrac{\rho}{4d}  |v_F \q|^2 } \ .
\end{align} \end{linenomath*}
This $n_s (\q)$ is a scaling function of $\tilde q = v_F |\q|/\Delta(0)$. The form of the scaling function depends on the dimensionality. In 2D we have
 \begin{linenomath*} \begin{align}
n_s^{2D}(\q)  \simeq \int_{-\infty}^{\infty}  dx \frac{1}{\frac{1}{4}\tilde q^2}  \times \left[\frac{1}{\sqrt{ 1 + x^2 }}  -  \frac{1}{\sqrt{ 1 + x^2  +\frac{1}{4}  \tilde q^2}} \right] \label{nsq} ,
\end{align} \end{linenomath*}
where we have approximated $\Delta(\omega) \simeq \Delta(0)$ and $\Lambda/\Delta(0) = \infty$, which holds to a  good numerical accuracy. Evaluating the frequency integral we find:
 \begin{linenomath*} \begin{align}
\label{nsqanalytic}
n_s^{2D}(\tilde q) =
  \frac{1}{\frac{1}{4}\tilde q^2} \log \left(1 + \frac{1}{4}\tilde q^2 \right).
\end{align} \end{linenomath*}
In 3D we have  \begin{linenomath*} \begin{align} \notag
&n_s^{3D}(\tilde q) = \frac{12}{\tilde q^3} \int dx \left[ \frac{{\tilde q}/{2}}{\sqrt{x^2 + 1} }- \arctan \left( \frac{{\tilde q}/{2}  }{\sqrt{x^2 + 1}} \right) \right] = \\
&\frac{12}{\tilde q^3 }\left[ \sqrt{4 + \tilde q^2 } \times \text{arctanh}\left( \frac{\tilde{q}}{\sqrt{4 + \tilde q^2}} \right)-\tilde q \right] \ . \label{nsqanalytic3d}
\end{align} \end{linenomath*}
We plot $n_s^{2D}(\tilde q)$ and $n_s^{3D}(\tilde q)$ in Fig.\  \ref{nsqFig}a. We see that both functions decrease with increasing ${\tilde q}$, i.e., with decreasing $\Delta_0$ for a given $\q$. At $f = f_c -0^+$,  $n_s^{2D,3D}(\tilde q)$ vanishes for any finite $\q$. {A suitably defined momentum-dependent compressibility $\kappa(\q) ~ \partial^2 \Pi_\Delta(\Omega,\q)/\partial \Omega^2\big|_{\Omega = 0} $ follows the same trend (Fig.\  \ref{nsqFig}b).}
  This behavior is indeed fully expected as  for $|v_F \q| \gg \Delta(0)$, the system is effectively in the normal state, where  the $U(1)$ symmetry is preserved and gauge (phase) fluctuations do not cost any energy.

\section*{\uppercase{Discussion}}
\label{concsec}

In this communication, we analyzed a $T=0$ superconductor-normal state transition for  a model of fermions coupled by a frequency-dependent interaction $V(\Omega)$, which has a repulsive constant part $f$ and an $\Omega$-dependent attractive part. For $f <0$, $V(\Omega)$ is fully attractive, and the system displays a conventional $s$-wave superconductivity with a sign-preserving $\Delta (\omega)$  along the Matsubara axis.  At
  $0<f <1$, the interaction $V(\Omega)$ is attractive at small frequencies, but repulsive at large $\Omega$.  In this case, superconductivity is still present at $T=0$, but the gap function has a node along the Matsubara axis, at some finite $\omega = \omega_0$. Because a nodal point of $\Delta (\omega)$ is a center of a dynamical vortex, the superconducting states at $f <0$ and at $f >0$ are topologically different.  We analyzed the topological transition at $f=0$ and argued that the vortex emerges at an infinite frequency at $f=0$ and moves to a finite $\omega_0$ at a finite $f$.  This is an expected behavior, consistent with earlier analysis of a similar model \cite{PhysRevB.104.L140501}. We also analyzed how the critical $f$ for such topological transition changes if we set a finite UV cutoff for the interaction.

  The superconducting state with a sign-changing $\Delta (\omega)$ persists also for $f >1$, when $V(\Omega)$ becomes positive  at all frequencies, and vanishes at a finite $f_c$.  The key part of our work is the analysis how the gap function $\Delta$ and the frequency $\omega_0$, where $\Delta$ changes sign, behave at $f \leq f_c$.

    We found that the gap function at zero frequency, $\Delta (0)$, vanishes exponentially fast with $(f_c - f)$. The frequency $\omega_0$ vanishes as well, but parametrically slower as $\sqrt{f_c - f}$. We argued that this
     parametrical difference between $\Delta (0)$ and $\omega_0$  allows one to obtain a non-zero  solution of the gap equation for all $f <f_c$.  We note that the transition at $f_c$ can be also interpreted from topological perspective, as the center of the dynamical vortex  reaches $\omega =0$ at $f=f_c$ and would have nowhere to go if superconductivity persisted above $f_c$.

We complimented the analysis of $\Delta (\omega)$ near $f_c$ by the analysis of the propagator of
 phase fluctuations. We have shown that the superfluid density and the compressibility, which control momentum and frequency parts of the propagator of a phase field, deviate from the BCS values at a generic $f <f_c$, but tend to the
  BCS values at $f \to f_c$ and undergo a finite jump at $f=f_c +0^+$  We showed that this, however, holds only for the superfluid density (and the compressibility) defined at strictly zero momentum.  We introduced a generic
    momentum-dependent $n_s (\q)$ and showed that it gradually vanishes at $f \to f_c$ for all $|v_F \q| > \Delta_0$. At $f$ slightly below $f_c$, this behavior holds or all $|\q|$ except the ones which are exponentially small in $f_c -f$.

There are multiple possibilities to extend our analysis. One extension, which we leave for further research, is a potential co-existence of even-frequency and odd-frequency superconducting orders at $f \leq f_c$. Such a state spontaneously breaks time-reversal symmetry.

Finally, our analysis is not constrained to electron-phonon interaction and is applicable to all cases when there is a near-constant repulsion and frequency-dependent, retarded attraction due to a boson exchange.
Other interesting candidates for a boson are exiton-polaritons in a microcavity \cite{PhysRevLett.104.106402, PhysRevB.93.054510} or cavity photons
\cite{PhysRevLett.122.133602}. Experimental cavity setups often come with a tuning knob which allows one to change the relative strength of repulsive and attractive components of the interaction (i.e., continuously change $f$ in our model). This  should allow one to observe phase transition at $f_c$ that we analyzed in this work.

\section*{\uppercase{Methods}}

\paragraph*{\bf{Derivation of the expression for the superfluid density by expanding the action in $\theta$.}}
\label{nsApp}

After the Hubbard-Stratonovich transformation is performed, the mixed boson-fermion action takes the form
(for the following derivation, compare, e.g., \cite{altland2010condensed}):

 \begin{linenomath*} \begin{align}
&\mathcal{S} = \mathcal{S}_\Delta - \int_{p,p^\prime} \bar{\Psi}_p G^{-1}(p,p^\prime) \Psi_{p^\prime}  \\ \notag
&\mathcal{S}_\Delta = \int_q \int d\omega d\omega^\prime \bar{\Delta}(\omega,q) V^{-1} (\omega - \omega^\prime) \Delta(\omega^\prime,q) \\ \notag
 &\Psi_p = \left(c_\uparrow(p), \bar{c}_\downarrow(-p)\right)^T, \quad p =  (\nu, \p) \\ \notag
  &G^{-1}(p,p^\prime) =  \notag   \begin{pmatrix} G_0^{-1}(p) \delta^{(d+1)}(p-p^\prime) & \Delta \left( \frac{1}{2}(\nu + \nu^\prime), p - p^\prime\right) \\[6pt] \bar{\Delta}\left( \frac{1}{2}(\nu + \nu^\prime), p^\prime - p\right) & - G_0^{-1} (-p) \delta^{(d+1)}(p -p^\prime) \end{pmatrix} \ .   \notag
\end{align} \end{linenomath*}
Here, the first argument of $\Delta$ contains the relative energy, and the second the total energy-momentum of the Cooper pair, compare Fig.\ \ref{pairingvariables}. Integrating out the fermions, we obtain
a purely bosonic action
 \begin{linenomath*} \begin{align}
\label{mainaction}
\mathcal{S} = \mathcal{S}_\Delta - \text{Tr} \ln ( - G^{-1}) \ ,
\end{align} \end{linenomath*}
where the trace runs over energy-momenta and Nambu indices. From Eq.\ \eqref{mainaction}, the gap equation is simply derived by setting $\delta\mathcal S /\delta \Delta(\omega) = 0$. We look for mean-field solutions which have zero total energy-momentum, i.e., contain a delta-function $\delta^{(d+1)} (q)$.

To find the action of the Goldstone-mode, we insert the expansion from Eq.\ \eqref{thetaapprox},
 \begin{linenomath*} \begin{align}
 \label{thetaapproxApp}
\Delta (\omega, q ) &\simeq  \Delta (\omega) \times \left[\delta^{(d+1)}(q)+ i\theta(q)\right]   \\
\bar{\Delta}(\omega, q) &\simeq \Delta (\omega) \times \left[\delta^{(d+1)}(q)- i\theta(-q)\right] \notag ,
\end{align} \end{linenomath*}
 into \eqref{mainaction}. The $ \mathcal{O}(\theta^0)$ contribution from the $\mathcal{S}_\Delta$ term has the form
  \begin{linenomath*} \begin{align}
 \int d\omega \Delta(\omega) V^{-1} (\omega - \omega^\prime) \Delta(\omega) \times \delta^{(d+1)} (0) \ .
 \end{align} \end{linenomath*}
 The term $ \delta^{(d+1)} (0)$ should be interpreted as volume factor. The $\mathcal{O}(\theta)$-contribution cancels, and the $\mathcal{O}(\theta^2)$-contribution reads
  \begin{linenomath*} \begin{align} \label{Stheta1}
 &S_\theta^{(a)} =   \int_q \theta(q) \theta(-q)
\left[ \int \frac{d\omega d\omega^\prime}{(2\pi)^2} \Delta(\omega) ( - V^{-1})(\omega - \omega^\prime) \Delta(\omega^\prime) \right]  \ .
\end{align} \end{linenomath*}
 To expand the $(\text{Tr} \ln)$-term, it is convenient to split
 \begin{linenomath*} \begin{align}
&G^{-1} (p,p^\prime) = G_0^{-1}(p,p^\prime) + X(p,p^\prime) \\  \notag
&G_0^{-1}(p,p^\prime) = \delta^{d+1}(p - p^\prime) \begin{pmatrix} i\nu - \xi_\p & \Delta(\nu) \\ \Delta(\nu) & i\nu - \xi_{-\p} \end{pmatrix}  \\  \notag
&X(p,p^\prime) = \begin{pmatrix} 0 & \Delta\left( \frac{\nu + \nu^\prime}{2} \right) i \theta(p - p^\prime) \\-  \Delta\left( \frac{\nu + \nu^\prime}{2} \right) i \theta(p - p^\prime) \end{pmatrix}
\end{align} \end{linenomath*}
Now, the trace of the logarithm can be expanded as
 \begin{linenomath*} \begin{align}
&\text{Tr} \ln\left(-G^{-1} \right) = \text{Tr}  \ln\left( - G_0^{-1} \left( \mathbbm{1} + G_0 \cdot X \right) \right) =  \\ &
\notag \text{Tr}  \ln \left( - G_0^{-1} \right) + \text{Tr}   \left( G_0 \cdot X \right ) - \frac{1}{2} \text{Tr}   \left( G_0 \cdot X \cdot G_0 \cdot X \right) \ .
\end{align} \end{linenomath*}
The first term is independent of $\theta$, and the second term $\mathcal{O}(\theta)$ cancels. To evaluate the third term $\mathcal{O}(\theta^2)$, it is convenient to introduce center-of-mass coordinates as $(\Omega, \q) = q = p - p^\prime$ and $(\omega, \k) = k = \frac{p + p^\prime}{2}$.
{In these coordinates,  $ - \frac{1}{2} \text{Tr} \left( G_0 \cdot X \cdot G_0 \cdot X \right)$
is \begin{widetext}
 \begin{linenomath*} \begin{align}
-\frac{1}{2} \int_{k,q} \text{tr} \left[ G_0(k + q/2) \cdot X(k + q/2, k - q/2) \cdot G_0(k - q/2) \cdot X(k - q/2, k + q/2) \right] ,
\end{align} \end{linenomath*}
\end{widetext}
where $\text{tr}$ acts in
the
spinor space. After straightforward algebra, the combination of this term
} and \eqref{Stheta1}
 yields
 $\mathcal{S}_\theta$ from the main text, Eq.\ \eqref{GSaction}.\\

\paragraph*{\bf{Derivation of the expression for the superfluid density from the particle-particle susceptibility.}}
\label{chiapp}

An alternative way of deriving the expressions for the superfluid density $n_s$ and dynamical compressibility $\kappa$ is by computing the full particle-particle susceptibility $\chi$ from Feynman diagrams. The basic building blocks for the diagrams are the normal and anomalous Green's functions,
 \begin{linenomath*} \begin{align}
\label{Gmeanfieldmain}
G_{\alpha\beta}(\omega, \k) &= - \delta_{\alpha\beta}
 \frac{i\omega + \xi_\p}{\omega^2 + \xi_\k^2 + \Delta(\omega)^2}\ ,  \\ \notag
    F_{\alpha\beta}(\omega, \k) &= i\sigma_{\alpha\beta}^y
     \frac{\Delta(\omega)}{\omega^2 + \xi_\k^2 + \Delta(\omega)^2} \ ,
  \end{align} \end{linenomath*}
where $\Delta(\omega)$ is {chosen as real},  $\alpha, \beta$
 are spin indices, and $\sigma^y$ is a Pauli matrix.

\begin{figure}
\centering
\includegraphics[width=.7\textwidth]{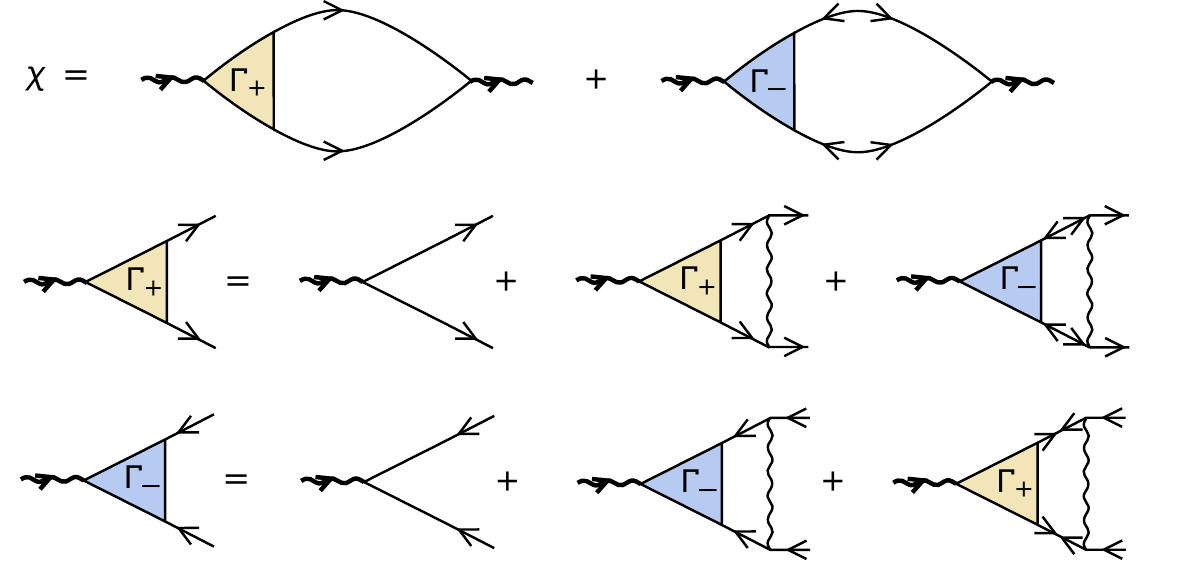}
\caption{\textbf{Diagrammatic representation of $\chi$}. Straight lines with a single arrow represent normal fermion propagators $G$, lines with a double arrow anomalous propagators $F$. Thin wavy lines represent interactions $V$, thick wavy lines the pairing field $\Delta$.}
\label{chidiags}
\end{figure}

The pairing susceptibility $\chi (q)$ can be represented as the sum of two contributions containing renormalized vertices $\Gamma_+, \Gamma_-$, see Fig.\ \ref{chidiags}. The vertices   satisfy
   the   two coupled Bethe-Salpeter equations
    and have poles corresponding to transverse (phase) fluctuations and longitudinal (Higgs) fluctuations (see e.g., Ref.\ \cite{lara}).
     One can verify that to describe only phase fluctuations one has to take      $\Gamma \equiv \Gamma_+ = - \Gamma_{-}$.  The single equation for $\Gamma$ then reads

  \begin{widetext}
 \begin{linenomath*} \begin{align}
\label{Gammaeq}
&\Gamma(\omega,q) = 1 + \int \frac{d\omega^\prime}{2\pi} \Gamma(\omega^\prime,q) V(\omega - \omega^\prime) \times A(\omega^\prime, q) ,  \\  \notag
& A(\omega^\prime, q) = \int \frac{d\k}{(2\pi)^d} \bigg[ G(\omega^\prime + \Omega/2, \k + \q/2) G(-\omega^\prime + \Omega/2, -\k + \q/2) \\ \notag  &+ F(\omega^\prime + \Omega/2, \k + \q/2) F(-\omega^\prime + \Omega/2, -\k + \q/2) \bigg] .
\end{align} \end{linenomath*}
\end{widetext}

 {Inserting
     the forms of $G$ and $F$, Eqs. \eqref{Gmeanfieldmain}, we find } that $A(\omega,q)$ and the modified particle-particle bubble $\Pi_\Delta(\omega,q)$, introduced in Eq.\ \eqref{Piqdefinition}, are related as
 \begin{linenomath*} \begin{align}
\label{PiArelation}
\Pi_\Delta(q) =
\int \frac{d\omega}{2\pi} A(\omega, q) \Delta^2(\omega) \ .
\end{align} \end{linenomath*}
To solve Eq.\ \eqref{Gammaeq}, we make an ansatz
 \begin{linenomath*} \begin{align}
\label{Gammaans}
\Gamma(\omega, q) = \frac{\Phi(\omega, q)}{c_1 |\q|^2 + c_2 \Omega^2} \ ,
\end{align} \end{linenomath*}
where $\Phi(\omega, q)$ is regular and non-vanishing for $q = 0$,
  and $c_1, c_2$ are some constants.
   Then, $\Gamma(\omega, 0) = \infty$, which formally solves \eqref{Gammaeq}.  To find the
    values of $c_1$ and $c_2$,
       we expand $A(\omega, q)$ from Eq.\ \eqref{Gammaeq} in $\Omega, |\q|$:
 \begin{linenomath*} \begin{align}
\label{Aexp}
A(\omega, q) \simeq A(\omega, 0) + a_{\q} (\omega) |\q|^2 + a_{\Omega} (\omega) \Omega^2 \ .
\end{align} \end{linenomath*}
Likewise, we expand
 \begin{linenomath*} \begin{align}
\label{Phiexp}
\Phi(\omega, q) \simeq \Phi(\omega, 0) + \phi_{\q}(\omega) |\q|^2 + \phi_{\Omega} (\omega) \Omega^2 \ .
\end{align} \end{linenomath*}
The coefficients $A(\omega, 0), a_{\q}, a_{\Omega}$ are known, while the coefficients $\Phi(\omega, 0), \phi_{\q}, \phi_{\Omega}, c_1, c_2$ are not known. We substitute Eqs.\ \eqref{Gammaans}, \eqref{Aexp}, \eqref{Phiexp} into
Eq.\ \eqref{Gammaeq}, which yields
\begin{widetext}
 \begin{linenomath*} \begin{align}
&\Phi(\omega, 0) + \phi_{\q} (\omega) |\q|^2 + \phi_{\Omega} (\omega) \Omega^2 =   c_1 |\q|^2 + c_2 \Omega^2 +\\  \notag &\int \frac{d\omega^\prime} {(2\pi)} V(\omega - \omega^\prime) \left [ \Phi(\omega^\prime, 0) + \phi_{\q} (\omega^\prime) |\q|^2 + \phi_{\Omega}(\omega^\prime) \Omega \right] \times \left[A(\omega^\prime, 0) + a_{\q} (\omega^\prime) |\q|^2 + a_{\Omega}(\omega^\prime) \Omega^2 \right]  \ .
\end{align} \end{linenomath*}
\end{widetext}
We now  compare the prefactors for $\mathcal{O}(1)$, $\mathcal{O}(|{\bf q}|^2)$, and $\mathcal{O}(\Omega^2)$  on both sides of this equation.   At order $\mathcal{O}(1)$, we have
 \begin{linenomath*} \begin{align}
\label{gapeqapp}
\Phi(\omega, 0) = \int \frac{d\omega^\prime}{2\pi} V(\omega - \omega^\prime) \Phi(\omega^\prime, 0) A(\omega^\prime, 0)\  .
\end{align} \end{linenomath*}
{
One can easily verify that the solution is
$\Phi(\omega, 0) = \Delta(\omega) \gamma$,
  where $\gamma$  is
  some constant, and $\Delta(\omega)$ is a solution of the gap equation}.
  Comparing the prefactors for the  $\mathcal{O}(|\q|^2)$-term, we get
 \begin{linenomath*} \begin{align}
\label{c1eq}
\phi_{\q}(\omega) = c_1 + \int \frac{d\omega^\prime}{2\pi} \phi_{\q} (\omega^\prime) A(\omega^\prime, 0) V(\omega - \omega^\prime) +  \int \frac{d\omega^\prime}{2\pi} \Phi(\omega^\prime, 0) a_{\q} (\omega^\prime) V(\omega - \omega^\prime) \ .
\end{align} \end{linenomath*}
Multiplying \eqref{c1eq} by $1/(2\pi) \times \Phi(\omega, 0) A(\omega, 0)$ and integrating over $\omega$, we obtain
 \begin{linenomath*} \begin{align} \notag
&\int \frac{d\omega}{2\pi} \phi_{\q} (\omega) \Phi(\omega, 0) A(\omega, 0) = \\   & c_1\! \int \frac{d\omega}{2\pi} \Phi(\omega, 0) A(\omega, 0) +  \int \frac{d\omega^\prime}{2\pi} \phi_{\q} (\omega^\prime) \Phi(\omega^\prime,0) A(\omega^\prime, 0) +  \int \frac{d \omega^\prime}{2\pi} \Phi^2(\omega^\prime, 0) a_{\q}(\omega^\prime)  ,
\label{cancelformula}
\end{align} \end{linenomath*}
where the gap equation in the form \eqref{gapeqapp} was applied twice on the right hand side of Eq.\ \eqref{cancelformula}. Cancelling the identical terms on both sides, we
 solve for $c_1$:
 \begin{linenomath*} \begin{align}
\label{b1result}
c_1 = - \frac{\gamma}{\int \frac{d\omega}{2\pi} \Delta(\omega) A(\omega, 0) } \times \int \frac{d\omega}{2\pi}  \Delta^2(\omega) a_{\q} (\omega) \  .
\end{align} \end{linenomath*}
In a similar fashion we obtain
 \begin{linenomath*} \begin{align}
\label{b2result}
c_2 = - \frac{\gamma}{\int \frac{d\omega}{2\pi} \Delta(\omega )A(\omega, 0) } \times \int \frac{d\omega}{2\pi} \Delta^2(\omega) a_{\Omega} (\omega) \ .
\end{align} \end{linenomath*}
Combining \eqref{b1result} , \eqref{b2result} and \eqref{Gammaans}, we
 obtain, to the leading order in $\Omega, |\q|$,
 \begin{linenomath*} \begin{align}
&\Gamma(\omega, q) \simeq \frac{\Phi(\omega, 0)}{c_1|\q|^2 + c_2 \Omega^2} = \frac{\Delta(\omega) \times \int \frac{d\omega^\prime}{2\pi} \Delta(\omega^\prime) A(\omega^\prime, 0)}{n_s |v_F\q|^2 + \kappa \Omega^2}
\end{align} \end{linenomath*}
where
 \begin{linenomath*} \begin{align}
& n_s = - \frac{1}{v_F^2}\int \frac{d\omega}{2\pi} \Delta(\omega^2) a_{\q}(\omega) \\
&\kappa = - \int \frac{d\omega}{2\pi} \Delta^2(\omega) a_{\Omega}(\omega) \ ,
\end{align} \end{linenomath*}
are the same as in the main text and in the previous section, see Eqs.\ \eqref{nsintegral}, \eqref{kappaintegral}.
Note that an arbitrary  constant $\gamma$ has cancelled out, as it indeed should.  The susceptibility $\chi$ has the same pole structure as $\Gamma$. To the leading order in $q$  we have
 \begin{linenomath*} \begin{align}
&\chi(q) \simeq \int \frac{d\omega}{2\pi} \Gamma(q,\omega) A(\omega,0)
= \frac{\left( \int \frac{d\omega}{2\pi} \Delta(\omega) A(\omega, 0) \right)^2}{n_s |v_F\q|^2 + \kappa \Omega^2}.
\end{align} \end{linenomath*}

\paragraph*{\bf{Data availability.}}
The numerical data used in the analysis in this work are available upon request from the corresponding author.

\paragraph*{\bf{Code availability.}}
The codes used to generate the numerical data  are available upon request from the corresponding author.

\paragraph*{\bf{Acknowledgment.}}We thank  Matthias Hecker, Dan Phan, and  Shang-Shun Zhang for useful discussions. The work was supported by NSF grant DMR-1834856.

\paragraph*{\bf{Competing interests.}} The authors declare no competing interests.

\paragraph*{\bf{Author contributions.}}
D.P. and A.V.C. performed the analytic calculations. D.P. performed the numerical computations. Both authors contributed to the discussion of results and to writing the manuscript.

\end{document}